\documentclass[reprint,superscriptaddress,preprintnumbers,amsmath,amssymb,aps,prd,longbibliography]{revtex4-2}

\usepackage{graphicx}
\usepackage{xcolor}
\usepackage[sort&compress]{natbib}
\usepackage{amsmath,amssymb,bm,bbm,slashed,subdepth}
\usepackage{xr-hyper}
\usepackage[colorlinks=true
,urlcolor=blue
,anchorcolor=blue
,citecolor=blue
,filecolor=blue
,linkcolor=red
,menucolor=blue
,linktocpage=true
,pdfproducer=medialab
,pdfa=true
]{hyperref}
\usepackage{cleveref}
\usepackage{enumerate}
\usepackage{subfigure}
\usepackage{setspace}
\usepackage{booktabs, tabularx}
\usepackage{units}
\usepackage{placeins}
\usepackage{multirow}
\usepackage{mathtools}

\begin{document}

\preprint{DESY-24-111}

\title{
Complete Gravitational-Wave Spectrum of the Sun
}

\author{Camilo Garc\'ia-Cely}
\affiliation{Instituto de F\'{i}sica Corpuscular (IFIC), Universitat de Val\`{e}ncia-CSIC, Parc Cient\'{i}fic UV, C/ Catedr\'{a}tico Jos\'{e} Beltr\'{a}n 2,
E-46980 Paterna, Spain}

\author{Andreas Ringwald}
\affiliation{Deutsches Elektronen-Synchrotron DESY, Notkestr. 85,
22607 Hamburg, Germany}

\begin{abstract}
The high-temperature plasma in the solar interior generates stochastic gravitational waves (GWs). Due to its significance as the primary source of high-frequency GWs in the solar system, we reexamine this phenomenon highlighting some physical processes, including the contribution of macroscopic hydrodynamic fluctuations. Our analysis builds upon several studies of axion emission from the Sun, particularly in relation to the treatment of plasma effects. 
The resulting GW spectrum is comparable to many well-motivated early Universe signals, yet orders of magnitude below the current sensitivities of axion helioscopes such as (Baby)IAXO.
\end{abstract}

\maketitle

\textbf{Introduction.} While current observational efforts~\cite{LIGOScientific:2017vwq,LIGOScientific:2016aoc,EPTA:2023fyk,Reardon:2023gzh,NANOGrav:2023gor,Xu:2023wog} to detect gravitational waves (GWs) have predominantly focused on frequencies below a few kHz,  a growing community is  seriously considering higher frequencies with the hope of detecting Early Universe signals~\cite{Aggarwal:2025noe}.  This prompts the question of identifying GW backgrounds associated with the Standard Model (SM) at this frequency range. The dominant source of such GWs 
is the Sun.

\begin{figure}[t]
\includegraphics[height=0.365\textheight]{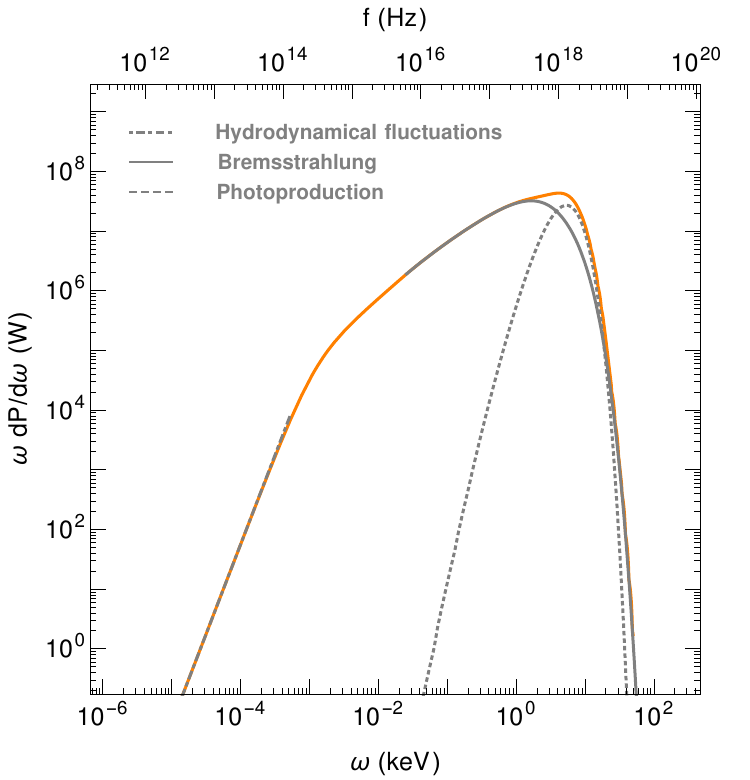}
\caption{ 
GW power per unit logarithmic frequency, originating from the solar thermal plasma as a function of the frequency, $\omega = 2\pi f$.  
}
\label{fig:GWspectrum}
\end{figure}

A seminal work by Weinberg~\cite{Weinberg:1965nx} estimated a total power  of approximately $6 \times \unit[10^{7}]{W}$. Although this figure turns out to be an excellent approximation (we find $1.3 \times \unit[10^{8}]{W}$), 
his spectrum was 
calculated only for soft gravitons from proton and electron collisions and neglected radial dependences --particularly of plasma effects~\cite{1954AuJPh...7..373S, Raffelt:1985nk}-- as well as additional contributions such as that of photoproduction~\cite{Voronov:1973kga, Galtsov:1974yu}. Employing a state-of-the-art solar model, we revisit the computation of the solar GW spectrum. 
For this we rely on several studies of axion emission from the Sun and show that they can be adapted to solar GWs, particularly those related to the treatment of plasma effects. 
Additionally, we point out the contribution from hydrodynamical fluctuations,  previously overlooked, but essential for completing the spectrum in its low-frequency range. 
Our prediction of the complete solar GW power spectrum
is summarized in Fig.~\ref{fig:GWspectrum}. 

We organize the discussion as follows. We start by reviewing properties of the solar plasma that will be relevant throughout, then we compile all the contributions to the GW spectrum of the Sun and clarify what has been omitted in the literature. Subsequently, we discuss implications for axion helioscopes.
In the Supplemental Material we provide the details of our calculations.  Throughout we adopt Heaviside units, take $\hbar=c=1$, and utilize the Minkowski metric $\eta_{\mu\nu} = \text{diag}(+ - - -)$.

\textbf{The solar plasma.}   We  model the Sun as a non-relativistic plasma composed of electrons ($e$) and nucleons ($Z$) following a Maxwell-Boltzmann distribution. We take the corresponding temperature and number density distributions from the B16-GS98 Solar model~\cite{Vinyoles:2016djt}, and estimate the electron density assuming full ionization. Elements heavier than Helium are negligible and are excluded from the analysis.  

GWs originate through microscopic and macroscopic mechanisms, respectively corresponding to  particle collisions emitting one graviton, $h$,~\footnote{The emission of multiple gravitons becomes relevant only at Planck-scale temperatures~\cite{Ghiglieri:2024ghm}} and GW emission sourced by hydrodynamic fluctuations.
The former corresponds  to frequencies much larger than that of collisions in the solar plasma, so that there is sufficient time for them not to interfere with each other~\cite{Weinberg:1972kfs}.
Below all collision frequencies of the plasma, fluctuations are large compared with the characteristic microscopic dimensions, allowing a hydrodynamical approach to calculate GW emission~\cite{Ghiglieri:2015nfa, landau9}. 

\begin{figure}[t]
\includegraphics[height=0.32\textheight]{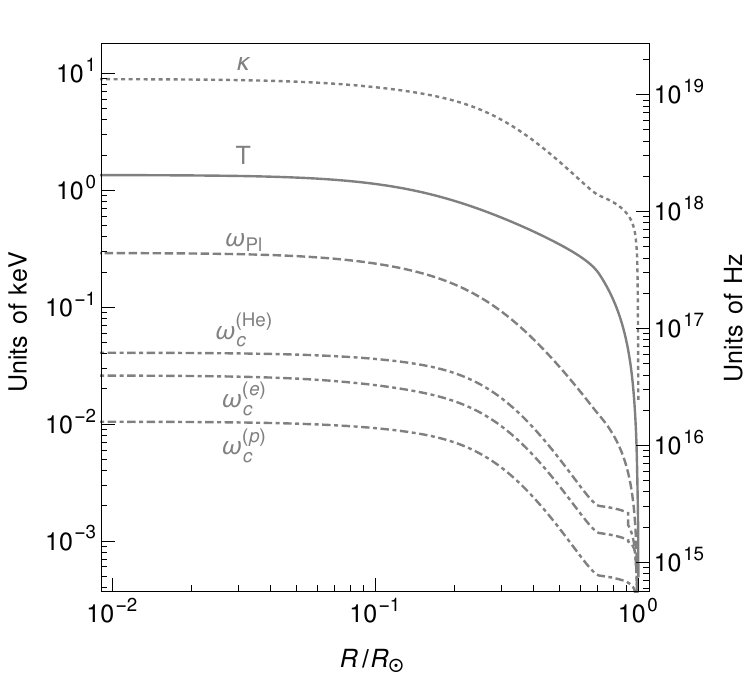}
\caption{ 
Screening scale (dotted), temperature (solid), plasma frequency (dashed), and collision frequencies (dot-dashed) as a function of distance from the center for the adopted solar model~\cite{Vinyoles:2016djt}. See text for details. 
}
\label{fig:figquantities}
\end{figure}

The estimation of the collision frequencies for each species, $\omega^{(i)}_c$,  is therefore of utmost importance for this work.
By following the approach detailed next, we obtain the results presented in Fig.~\ref{fig:figquantities}. We first calculate the viscosity cross section, $\sigma_V$.
For instance, for $eZ$, 
\begin{equation}
\sigma_V^{(eZ)}=\int  \frac{\mathrm{d}\sigma}{\mathrm{d} \Omega}^{(eZ)}
 \sin^2\theta \mathrm{d} \Omega \simeq 
 \frac{2 \pi \alpha^2 Z^2}{E_i^2}L_{\text{Coulomb}} \,,
\label{eq:sigmav}
\end{equation}
and similarly for $ee$.  The collision frequencies are then estimated from the corresponding scattering rate as $\omega^{(e)}_c = n_e  \sigma_V^{(ee)}v + \sum_Z n_Z  \sigma_V^{(eZ) } v$. 

To define $L_{\text{Coulomb}}$, we note the following. Screening gives  rise, in the non-relativistic limit,
 to an effective  Yukawa potential  with a range given by the Debye-H\"uckel screening scale,   
$\kappa=\left(4 \pi \alpha(n_e+\sum_Z Z n_Z)/T \right)^{1/2}$~\cite{1954AuJPh...7..373S}, see Fig.~\ref{fig:figquantities}.    For instance, for $eZ$ collisions, $V_{\text{Coulomb}} = -Z \alpha/r \to - Z \alpha e^{-\kappa r}/r $. 
For non-relativistic collisions, the scattering rate associated with these potentials may be non-perturbative. Nevertheless, in the Sun the scattering largely occurs either in the Born regime, where perturbation theory is applicable, or in the semi-classical regime, where classical physics can be employed. 
In these cases, the viscosity cross section is given by the last equality in Eq.~\eqref{eq:sigmav}, where the Coulomb logarithm, $L_{\text{Coulomb}}$,
 equals $\log 2p_i/\kappa$ and $ \log E_i/(Z^2\alpha \kappa)$, in the Born and semi-classical approximations, respectively~\cite{Weinberg:2019mai, Khrapak:2003kjw}. Throughout,   $p_i$ ($p_f$) denotes the initial (final) momentum of the colliding particles in the center-of-mass frame, and $E_i$ is their total kinetic energy.
 
 We note that the collision frequencies should be regarded as transitions scales, particularly because Eq.~\eqref{eq:sigmav} is valid only up to logarithmic accuracy~\cite{pitaevskii2012physical}  and different prescriptions exist for the velocity averaging~\cite{Weinberg:2019mai}. Here, we follow~\cite{Weinberg:1972kfs} and take $v$ equal to its thermal average. For further details, see Supplemental Material.

\textbf{GWs from hydrodynamical fluctuations.} We closely follow Ref.~\cite{Ghiglieri:2015nfa} to calculate this contribution.  In essence, GWs in this limit are sourced by tensor fluctuations of the energy-momentum tensor, whose Fourier transform is largely independent of the frequency and wave number.  
Being purely hydrodynamical, the tensor fluctuations are simply proportional to the shear viscosity, $\eta$, of the plasma. Plugging them in Einstein's Equations, one obtains 
\begin{equation}
\frac{\mathrm{d}P}{\mathrm{d}\omega}\Bigg|_\mathrm{Hydrodynamics}= \frac{16 G \omega^2}{\pi} \int_\text{Sun} \!\!\!\!\mathrm{d}^3 \mathbf{r} \,   \eta T\,.
\label{eq:hydro}
\end{equation}
While this emission has not been previously investigated for the Sun or stellar plasmas, it is known that it contributes to the GW emission from the  Early Universe plasma, that is, to the Cosmic Gravitational Microwave Background (CGMB)~\cite{Ghiglieri:2015nfa,Ghiglieri:2020mhm,Ringwald:2020ist}. The only conceptual difference here is that viscosity must be computed differently because the solar plasma is non-relativistic.  

Due to the significant mass ratio between protons and electrons, the primary source of shear viscosity may be attributed to momentum transfer involving protons. For this we note that proton-proton collisions in the Sun are semi-classical. Moreover, the  proton component of the solar plasma  is not strongly correlated, meaning that the average Coulomb potential energy is much smaller than the thermal energy (which follows from $\alpha\, n_p^{1/3}/T \sim 0.03-0.07 \ll1$). This justifies to employ the numerical fits to the  state-of-the-art simulations reported by Ref.~\cite{PhysRevE.90.033105} --more precisely, its Eq.~(31)-- for the calculation of the viscosity. 
In practice, this is just a slight modification of the Landau-Spitzer theory, which predicts $\eta \sim m_p v_p/\sigma^{(pp)}_V$~\cite{pitaevskii2012physical}.  
Let us remark that Eq.~\eqref{eq:hydro} loses its validity  at very low frequencies --where large-scale solar phenomena must contribute to the GW signal~\footnote{High-Reynolds-number turbulence from stellar convection has been claimed to contribute at micro-Hertz scales a GW power $\omega {\rm d}P/{\rm d}\omega \approx 4\times 10^{-12}\,{\rm W}$ ~\cite{Bennett:2014tba}}-- or when $\omega\gtrsim\omega^{(p)}_c$ as sound waves associated with the hydrodynamical fluctuations are damped by viscosity~\cite{Klose:2022knn}.

The solar GW power spectrum from hydrodynamic fluctuations is given by the dashed-dotted line in Fig.~\ref{fig:GWspectrum}, where we conservatively put a cut-off for frequencies above  $0.05\,\omega_c^{(p)} \big|_{\rm Center}$ in light of damping.


\textbf{GWs from particle collisions.} This contribution to the solar GW spectrum is obtained by thermally averaging  individual graviton emission rates, 
\begin{equation}
\frac{\mathrm{d}P}{\mathrm{d}\omega}\Bigg|_\mathrm{Collisions}= \int_\text{Sun} \!\!\!\!\mathrm{d}^3 \mathbf{r} \,  \sum_{ i }  \omega \left\langle \frac{\mathrm{d}\Gamma^{(i)}(\mathbf{r})  }{\mathrm{d}\omega\mathrm{d}V} \right\rangle \,.
\label{eq:collisions}
\end{equation} 
Here $i$ runs over the processes detailed below and whose rate, to leading order in the screening scale, are summarized in Table 1 of the Supplemental Material.
These rates are calculated in the traceless-transverse gauge~\footnote{At these frequencies, the Sun is effectively in free fall and thus at rest in this gauge, resulting in no medium response, see e.g.~\cite{Ratzinger:2024spd}.} by squaring the relevant amplitudes using CalcHEP~\cite{Belyaev:2012qa} and 
FeynRules~\cite{Alloul:2013bka} and taking the non-relativistic limit for electrons and nucleons (see also~\cite{Choi:1994ax,Donoghue:1994dn,Patel:2016fam}).  We assume spherical symmetry for the integration in the Sun.

\emph{Bremsstrahlung ($eZ\to eZ h $ and  $ee\to ee h $).}  We account for plasma effects following the bremsstrahlung treatment of Ref.~\cite{Raffelt:1985nk}. This entails modifying the propagator of the photon exchanged by the charged particles  with an effective mass equal to $\kappa$, as suggested by the aforementioned Yukawa potential. 
In addition to  corrections of order $\kappa^2/p_i^2$, this introduces an infrared cut-off. This is essential because, in the absence of screening,  the rates exhibit a logarithmic divergence for soft gravitons.
A more rigorous formulation can be achieved using thermal field theory, wherein the inclusion of hard thermal loop self-energies in the propagators naturally regulates such divergences across a broad frequency range, as has been done for axions, see~\cite{Altherr:1992mf}. In this formalism, the screening scale squared $\kappa^2$ appears as a photon self-energy in the static limit ($\omega\to0$)~\cite{Kapusta:1989tk, Raffelt:1996wa}.

Notably, our explicit computations  reveal that the bremsstrahlung rates depend on a common regulating logarithm,  
which for soft gravitons approaches the one entering the viscosity cross section of Eq.~\eqref{eq:sigmav}. This is not a coincidence,   soft theorems~\cite{Weinberg:1965nx} arising from basic principles of quantum field theory dictate that  
$\omega \mathrm{d}\Gamma^{(eZ)}/\mathrm{d}\omega \mathrm{d}V |_\mathrm{soft} =(32 G / 5 \pi) E_i^2 \sin ^2 \theta \,  n_e n_Z   \sigma^{(eZ)}_V v$ for $\omega \ll E_i \sim T$, 
and a similar expression for $ee$ with a factor $1/2$~\footnote{Although Refs.~\cite{Weinberg:1965nx, Weinberg:1972kfs} seemingly omitted it, this symmetrization factor is explicitly included in the $pp$ chain~\cite{Bethe:1938yy} and in axion bremsstrahlung from $ee$~\cite{Raffelt:1985nk}.
}.  
Applying this to the center of the Sun 
is how Weinberg~\cite{Weinberg:1965nx,Weinberg:1972kfs}  originally found
a flat power  spectrum, $\mathrm{d}P/\mathrm{d}\omega$, with only a   minor logarithmic dependence. Interestingly, despite its quantum-mechanical origin,   this formula coincides~\cite{Weinberg:1972kfs,Gould:1985, Steane:2023gme} with the classical emission rate of GWs associated with  the quadrupole of the $eZ/ee$  hyperbolic orbit  for sufficiently soft radiation \footnote{Note that this is in agreement with the arguments presented in \cite{Carney:2023nzz} concerning the treatment of a classical GW as a collection of individual gravitons (see also~\cite{Dyson:2013hbl}).}. Beyond the soft limit, ours and the classical rate~\cite{1967PhRv..158.1243C} differ, with the latter overestimating the spectrum because collisions in the Sun's interior are outside the classical regime~\cite{Gould:1985}, see Supplemental Material.

Weinberg's calculation motivated studies of bremsstrahlung beyond the soft limit (See also ~\cite{Weinberg:2019mai, Pradler:2021ppc,Flauger:2019cam} ). 
Reference~\cite{Barker:1969jk, Papini:1977fm} reported rates associated with $eZ$ --but not $ee$-- using similar methods to ours but assuming an infinitely heavy nucleon and no screening.  Likewise, bremsstrahlung rates neglecting screening were calculated assuming that the emission comes from one electron moving in the Coulomb potential of a nucleon~\cite{1974SvPhJ..17.1713G, Gould:1981an} or another electron~\cite{Gould:1985}.  Our results 
agree with these studies in the appropriate limit~\footnote{A recent attempt to interpolate between the external field approximation and the semi-classical approach has been given in Ref.~\cite{Steane:2023gme}.  Our results with no screening do not match theirs as they neglect electrons' indistinguishability.}. 

It is worth mentioning that Ref.~\cite{Gould:1985}  calculated $ee$ bremsstrahlung rates in the non-relativistic limit through the following clever insight: \emph{photon}  bremsstrahlung from $ee$ results from their quadrupole, as their dipole vanishes by their indistinguishability, which allows adapting the $ee$ electromagnetic emission to gravitational radiation, with the adjustment being a mere global factor. Our rates, obtained through a different approach, support this conclusion. 

Finally, let us remark that our procedure relies on the validity of the Born approximation, which for $ee$ and $eZ$ holds throughout the Sun, except in the outermost regions, approximately the last 15\% of the solar radius. Since bremsstrahlung is
dominated by processes at the center, neglecting non-perturbative corrections to the graviton emission is justified.  The solar GW power spectrum from bremsstrahlung is given by the solid line in Fig.~\ref{fig:GWspectrum},  where we put a cut-off for frequencies below  $\omega_c^{(e)} \big|_{\rm Center}$. 
 We remark that for this we use the full rates reported in the Supplemental Material, where further details are provided. 

\emph{Photoproduction ( $\gamma Z/e\to  Z/e h$).} This is analogous to the Primakoff and Compton production of axions. Our results  with $\kappa=\omega_{\rm Pl}=0$ coincide with the differential rates for graviton emission reported in Ref.~\cite{Voronov:1973kga}.  
 Nevertheless, from this it is impossible to make sense of the integrated rate, as it diverges due to the collinear divergence appearing at $\theta=0$.   To tackle this problem, we again employ insights from axion physics~\cite{Raffelt:1985nk}. Concretely, we note  that charged particles in the solar plasma mutually interact through their Coulomb fields, leading to a pair correlation function~\cite{Landau:1980mil,Raffelt:1985nk,Raffelt:1987np}. 
This gives rise to an effective form factor, $F(\theta)$, that depends on the screening scale $\kappa$ and regularizes the total rate.

If one neglects such correlations and instead considers photons scattering off individual screened charges, one finds a rate with $|F(\theta)| \to |F(\theta)|^2$~\cite{DeLogi:1977qe}. In practice,  this corresponds to modifying the photon propagator with an effective mass, a picture which is appropriate if the scattering process is so slow that the charged particles can move around and rearrange themselves~\cite{Raffelt:1996wa}. 
This issue has been revisited recently for the Primakoff process of axions, accounting for the velocity in the form factor~\cite{Hoof:2021mld}.  From this study the difference between both approaches is expected to be negligible. See Supplemental Material for a detailed discussion.

Note also that the dispersion relation of the initial-state photon is influenced by the surrounding plasma, which imparts an effective mass to transverse photons equal to the plasma frequency, $\omega_\mathrm{Pl} = \left(4\pi n_e \alpha/m_e\right)^{1/2}$;  temperature-dependent corrections are negligible, as the plasma is non-relativistic and non-degenerate. Moreover, since the solar temperature is much greater than $\omega_\mathrm{Pl}$ (see Fig.~\ref{fig:figquantities}), the kinetic energies of all particles are well above the plasma frequency, and the GW spectrum remains largely insensitive to the photon's effective mass, a fact we have verified explicitly. The main consequence is the phase-space factor $p_i/\omega$, that provides a lower cut-off to the photoproduction contribution at each point of the Sun. 
We note that this is completely analogous to the Primakoff effect for axions, which we discuss in the Supplemental Material for completeness. It is noteworthy that Ref.~\cite{Galtsov:1974yu} calculated photoproduction rates using semi-classical methods and including the effect of the plasma frequency, but did not account for screening, which is the dominant effect.

The resulting solar GW power spectrum from photoproduction is given by the dashed line in Fig.~\ref{fig:GWspectrum}.

\textbf{Transition regime and other processes.}  As it shifts from macroscopic to microscopic fluctuations, $\mathrm{d}P/\mathrm{d}\omega$ transitions from $\propto\omega^2 $ to $ \propto\omega^0 $ within the range $ \omega^{(p)}_c \lesssim \omega \lesssim \omega^{(e)}_c $. Similar to the case of the CGMB, a thorough examination of this regime necessitates a hydrodynamical simulation. Nevertheless,   
we postulate a double-broken power law,  $ \omega^2/(\omega^2+\omega_0^2)$. 
The Landau-Spitzer viscosity for protons give  $\omega_0\approx ( \omega^{(p)}_c \omega^{(e)}_c )^{1/2}\propto 1/\eta$.  Notably, the interpolating law with the predicted  scale $\omega_0\propto 1/\eta$ is also found in analytical studies of the energy-momentum tensor for relativistic plasmas in the so-called shear channel~\cite{Ghiglieri:2015nfa}. It is also worth mentioning that  power shifts from $\mathrm{d}P/\mathrm{d}\omega \propto \eta \sim 1/\sigma_V $ to soft bremsstrahlung, where $\mathrm{d}P/\mathrm{d}\omega \propto  \sigma_V $.  Interestingly,  the same behavior is found for the CGMB because $\sigma_V\sim\alpha^2/T^2$.

Additionally, we find that bremsstrahlung --also known as free-free emission-- largely dominates over free-bound transitions —recombination of unbound charged particles into atomic states and a graviton— and bound-bound transitions —graviton emission from atomic transitions. In contrast, such processes are important for axions~\cite{Redondo:2013wwa,Pospelov:2008jk}, giving the axion solar spectrum distinct small peaks depending on the solar model~\cite{Hoof:2021mld}. While a similar effect might be expected for GWs due to the analogy with axions, the quadrupolar nature of graviton emission renders free-bound and bound-bound contributions negligible, as detailed in Supplemental Material III, following~\cite{Boughn:2006st, Rothman:2006fp, berestetskii1982quantum,2025SoPh..300....3B}. Hence, the GW spectrum from the Sun lacks such small peaks.

The same applies to photoproduction from longitudinal photons at frequencies $\omega_\mathrm{Pl}$, particularly in the presence of the solar magnetic field~\cite{Caputo:2020quz,OHare:2020wum,Guarini:2020hps, Hoof:2021mld}.
This follows from the fact that gravitons couple uniquely to photons and fermions, leading to fixed relative contributions in the solar GW spectrum. In contrast, axions possess model-dependent couplings to photons and fermions, allowing scenarios where bremsstrahlung is subdominant compared to such photoproduction  processes.  In the light of this,  the latter hardly contribute to the GW spectrum around $\omega_\mathrm{Pl}\sim\unit[0.3]{keV}$ or lower where bremsstrahlung dominates by more than five orders of magnitude.

\begin{figure}[t]
\includegraphics[height =0.35\textheight]{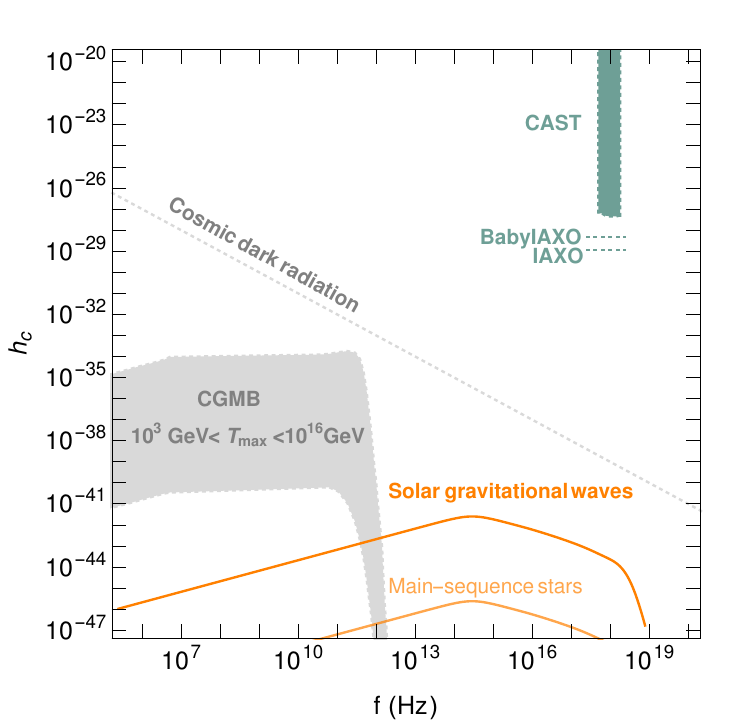}
\caption{
Characteristic GW strain versus frequency. Exclusion bound and corresponding projected sensitivities of current (CAST) and future (BabyIAXO, IAXO) axion helioscopes, together with the corresponding predictions from the solar GW spectrum in Fig.~\ref{fig:GWspectrum}, an estimate of the stochastic background from the main-sequence stars in the Milky Way, the CGMB spectrum, and the bound from the total radiation in the Universe. 
}
\label{fig:strain}
\end{figure}

\textbf{Discussion and outlook.}  
Our interpolation of the solar GW power spectrum including all processes is presented in Fig.~\ref{fig:GWspectrum}.  In terms of power, the relative contribution of each process is as follows: 25\% for photoproduction, 33\% for $ee$ bremsstrahlung, 24\% for $ep$ bremsstrahlung, 17\% for $eHe$ bremsstrahlung, with the rest attributed to the hydrodynamical contribution.  While the spectrum over several decades in frequency was not studied before, corrections of a few  to the above percentages --and correspondingly to the overall spectrum-- are expected, especially after accounting for degeneracy and effects beyond the Born approximation, as for solar axions~\cite{Hoof:2021mld}. 

Axion helioscopes~\cite{Sikivie:1983ip} can be used as detectors of  gravitational radiation exploiting the inverse Gertsenshtein effect~\cite{Gertsenshtein,Boccaletti1970}, by which GWs convert to electromagnetic waves in the presence of an external magnetic field.   Reference~\cite{Ejlli:2019bqj} presented  an exclusion bound from CAST~\cite{CAST:2017uph}, sensitivity prospects for IAXO~\cite{IAXO:2019mpb},  
and compared them against Weinberg's prediction of the solar GW spectrum. In Fig.~\ref{fig:strain} we present their findings in terms of the GW strain, $h_c\equiv 2( G\, \mathrm{d}P/\mathrm{d}\omega)^{1 / 2}/d_{\rm Sun}\omega^{1/2}$, together with the complete spectrum as derived in Fig.~\ref{fig:GWspectrum}. For comparison, we also display the CGMB, whose overall normalization depends upon the (unknown) maximum temperature of the primordial plasma~\cite{Ringwald:2020ist}.  
In addition, we show the cosmic dark radiation bound, arising from the fact that GWs contribute to the energy budget of the Universe in the form of radiation and are thus limited by Big Bang Nucleosynthesis (BBN) and Cosmic Microwave Background (CMB) observations~\cite{Aggarwal:2020olq,Aggarwal:2025noe}. 
In addition, we also derive prospects for BabyIAXO following Refs.~\cite{Ringwald:2020ist,IAXO:2019mpb}. The main challenge for GW detection in helioscopes lies in technological limitations related to magnetic fields and scalability. The strain sensitivity of a typical  helioscope  scales as $h_c \propto BLA^{1/2}$,  where $B$ is its magnetic field, $L$ is its length, and $A$ is its cross section. IAXO achieves $BLA^{1/2} \approx 90\,\mathrm{T\,m^2}$~\cite{Ringwald:2020ist}, yielding $h_c \sim 10^{-29}$ at $10^{18}\,\mathrm{Hz}$, still $\sim 15$ orders above the Sun spectrum, see Fig.~\ref{fig:strain}.
If these challenges are eventually overcome and the spectrum is measured with sufficient spatial and spectral resolution, this could provide a valuable window into the inner structure of the Sun, much like what has been envisioned for axions (see e.g.~\cite{Hoof:2023jol}).

For further comparison, we include an estimate of the stochastic background associated with main-sequence stars in the galaxy. Noting that these stars have masses similar to the Sun~\cite{Bovy_2017}, we approximate their GW spectrum using the one of the Sun, leading to $ h_{c\,\text{stars}} = \sqrt{\int dV \, n_{\text{stars}} \, d^2_{\rm Sun}/d^2} \, h_{c\,\text{Sun}}$. A similar estimation has been performed in Ref.~\cite{Brocato:1997tu} for stellar neutrinos. Adapting their results for the $^{7}\mathrm{Be}$ line, we find  $h_{c\, \text{stars}} \sim 10^{-4} h_{c\, \text{Sun}}$.    
 We note that this figure should be regarded as an order-of-magnitude estimate.
Our calculation of solar GWs reveals a clear similarity between gravitons and axions. Expanding on this idea, we could consider the emission of massive spin-2 particles from the Sun, which could also constitute the dark matter of the Universe, see e.g.~\cite{Babichev:2016hir,Babichev:2016bxi,Chu:2017msm}. 
Extending the methods discussed here to such particles, we find that the resulting spectrum is nearly identical to that of the solar GWs, differing only by an overall factor associated with the coupling of spin-2 particles to SM fields. In a separate publication~\cite{Future}, we will discuss this synergy and present observational prospects at helioscopes.

Our study represents a step towards establishing a multi-frequency GW background from SM processes, similar to that of neutrinos~\cite{Vitagliano:2017odj,Vitagliano:2019yzm}.  In fact, our approach can be extended to other stellar plasmas~\footnote{See~\cite{Gould:1985} for a discussion on white dwarfs and neutron stars} and to the early stages of the Universe after nucleosynthesis, where plasmas are also non-relativistic. Whether
we can aspire to probe such strain sensitivities in the
future remains an open question.

{\it Acknowledgements.}
%
We thank Diego Blas, Andrea Caputo, Valerie Domcke,  Javier Gal\'an, Maurizio Giannotti, Igor Irastorza, Mikko Laine, Jamie McDonald, Maxim Pospelov and Georg Raffelt for discussions. 
CGC is supported by a Ramón y Cajal contract with Ref.~RYC2020-029248-I, the Spanish National Grant PID2022-137268NA-C55 and Generalitat Valenciana through the grant CIPROM/22/69. 
A.R. acknowledges support by the Deutsche Forschungsgemeinschaft (DFG, German Research Foundation) under Germany’s Excellence Strategy - EXC 2121 Quantum Universe - 390833306 and under - 491245950.  
This article is based upon work from COST Action COSMIC WISPers CA21106, supported by COST (European Cooperation in Science and Technology).

\bibliographystyle{utphys-modified}
\bibliography{ref}

\clearpage
\onecolumngrid

\twocolumngrid
\setcounter{equation}{0}
\setcounter{figure}{0}
\setcounter{table}{0}
\setcounter{section}{0}
\setcounter{page}{1}
\makeatletter
\renewcommand{\theequation}{S\arabic{equation}}
\renewcommand{\thefigure}{S\arabic{figure}}
\renewcommand{\thetable}{S\arabic{table}}

\onecolumngrid

\begin{center}
   \textbf{\large SUPPLEMENTAL MATERIAL \\[.1cm] ``Complete Gravitational-Wave Spectrum of the Sun''}\\[.2cm]
  \vspace{0.05in}
  {Camilo Garc\'ia-Cely, and Andreas Ringwald}
\end{center}

\section{Viscosity cross sections in the Born and semi-classical regimes}

\begin{figure}[b]
\includegraphics[height=0.34\textheight]{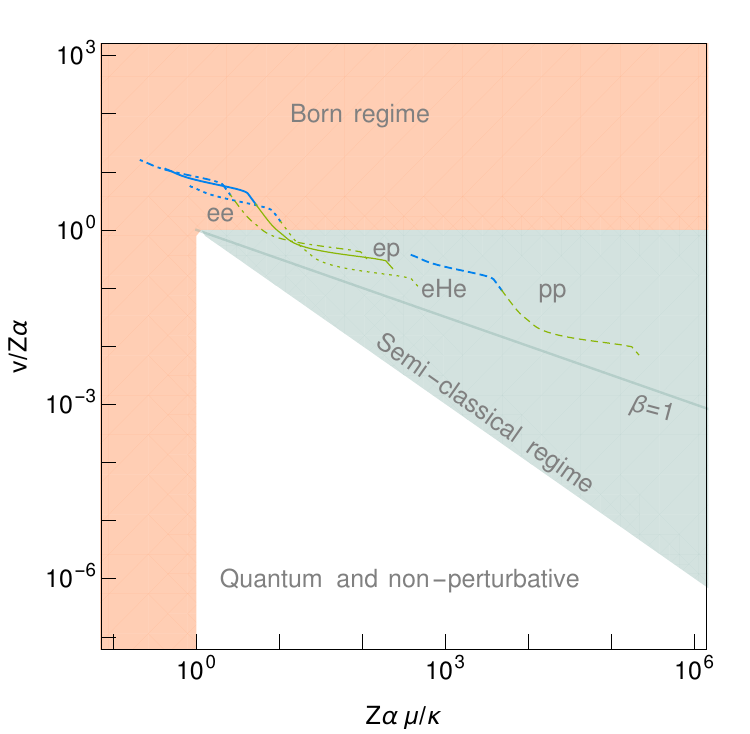}
\caption{ 
Scattering regimes of the indicated solar collisions due to the screened Coulomb potential. Perturbation theory is applicable in the Born regime, while classical physics is valid in the semi-classical regime. Different points correspond to various distances from the center of the Sun. See the text for details. 
}
\label{fig:fig1b}
\end{figure}

Viscosity cross sections can be used to estimate collision frequencies, and as discussed in the main text, Weinberg's soft theorem relates them to graviton bremsstrahlung. Hence, they are of utmost importance for this work. They are defined by
\begin{equation}
\sigma_V^{(ab)}=\int  \frac{\mathrm{d}\sigma}{\mathrm{d} \Omega}^{(ab)} \sin^2\theta \, \mathrm{d} \Omega \,.
\label{eq:sigmavAPP}
\end{equation}
For a Coulomb potential, $V(r)= -Z_a Z_b \alpha/r$, between two particles of charge $Z_a$ and $Z_b$ (in units of $e$), this cross section logarithmically diverges, because the corresponding $\mathrm{d}\sigma^{(ab)}/{\mathrm{d}\Omega}$ scales as $1/\sin^4(\theta/2)$. 
This is not really a problem because the Coulomb potential is screened. In the non-relativistic limit, this is described by an effective  Yukawa potential  with a range given by the Debye-H\"uckel screening scale,
\begin{align}
V(r)= -\frac{Z_a Z_b \alpha }{r}  e^{-\kappa r}\,, &&\text{and}&& \kappa^2=\frac{4 \pi \alpha}{T}\left(n_e+\sum_Z Z n_Z\right) \,.
\label{eq:potentialkappa}
\end{align}
Determining $\mathrm{d}\sigma^{(ab)}/{\mathrm{d}\Omega}$  for non-relativistic particles interacting by means of such a potential is non-trivial due to non-perturbative effects. Nevertheless, as we show in the sequel, for the relevant collisions in the Sun the scattering largely occurs either in the Born regime, where perturbation theory is applicable, or in the weakly coupled semi-classical regime, where classical physics can be employed.   In both scenarios, one obtains
\begin{equation}
\sigma_V^{(ab)}\simeq\frac{2 \pi \alpha^2 Z_a^2 Z_b^2}{E_i^2}L_{\text{Coulomb}}\,,
\label{eq:approx}
\end{equation}
where $L_{\text{Coulomb}}$ is the Coulomb logarithm. We emphasize this is valid to logarithmic accuracy, meaning that this expressions neglects quantities smaller than $L_{\text{Coulomb}}$. For a comprehensive discussion see e.g. Ref.~\cite{pitaevskii2012physical}. 
The Born regime applies for large velocities, namely $Z\alpha \ll v$, or for sufficiently small couplings, concretely $Z \alpha \ll \kappa/\mu$~\cite{pitaevskii2012physical}, where $\mu$ is the reduced mass. Ordinary quantum perturbation theory employing the potential in Eq.~\eqref{eq:potentialkappa} gives~\footnote{
Although this agrees with~\cite{Flauger:2019cam}, the viscosity cross section reported by Weinberg in his textbook~\cite{Weinberg:1972kfs} is $1/2$ of this.}
\begin{equation}
L_{\text{Coulomb}}=\log \left(\frac{2\mu v}{\kappa}\right)\,.
\label{eq:appBorn}
\end{equation}
The semi-classical regime takes place when the de Broglie wavelength is much smaller than the range of the potential, that is, $ \kappa \ll \mu v $ ~\cite{pitaevskii2012physical}. 
In this case there is a critical parameter, $\beta = |Z_a| |Z_b|  \kappa\alpha/\mu v^2 $, determining different sub-regimes~\cite{Khrapak:2003kjw}. For $\beta \ll 1$, of interest in this work, the plasma is weakly-coupled and the Coulomb logarithm is given by~\cite{pitaevskii2012physical,Khrapak:2003kjw}
\begin{equation}
L_{\text{Coulomb}}=\log \left(\frac{\mu v^2}{|Z_a| |Z_b|\alpha \kappa}\right) \,.
\label{eq:Lcclassical}
\end{equation}
The range of applicability of either regime along the position in the Sun is sketched  in Fig.~\ref{fig:fig1b}. 
The green portion of the lines corresponds to the 15\% outermost part, while the blue part corresponds to the  innermost 85\%. For simplicity we only show $ee$, $ep$, $eHe$ and $pp$ collisions. We also show the line $\beta=1$, above which the plasma is weakly coupled and Eq.~\eqref{eq:Lcclassical} can be used. Note that pp collisions are semi-classical, as mentioned in the main text.

\section{Graviton emission rates and Feynman rules}

\begin{table*}[t]
    \centering
    \begin{tabular}{|c|c|c|}
    \cline{1-2}
\multirow{2}{*}{\bf{Collision}} &
\multirow{2}{*}{ $\frac{\mathrm{d}\Gamma}{\mathrm{d}\omega \mathrm{d}V}$
}
\\&
\\   \cline{1-3}
{\small Photoproduction}
      &
\multirow{2}{*}{$ n_\gamma n_Z G Z^2  \alpha \pi\,  \delta(\omega-E_i) 
\frac{p_i}{\omega}
\int \mathrm{d}\!\cos\theta \left(\cot^2\!\frac{\theta}{2} [1+\cos^2\theta] |F(\theta)|^2 + {\cal O} (\omega_{\rm Pl}^2/\omega^2) \right)$}
&
\multirow{2}{*}{$|F(\theta)|^2 =\frac{\left(2 \omega  \sin \!\frac{\theta }{2}\right)^2}{\kappa ^2+\left(2 \omega  \sin \!\frac{\theta }{2}\right)^2} $}\\
 $\gamma\, Z/e\to  Z/e \,h$
&& \\\hline
{\small Bremsstrahlung}
       &
\multirow{2}{*}{     $
\frac{32 n_e n_Z  G  Z^2 \alpha^2 p_i }{15\omega} \left(\frac{1}{m_e}+\frac{1}{m_Z} \right) \left(3 (1+\xi ^2)
{L}
+10 \xi +{\cal O}(\xi_s^2)
\right)
  $ 
  }
&
$\xi =p_f/p_i$,\quad$\xi_s = \kappa/p_i$
\\
 $eZ\to eZ \,h $
&&
$\omega=E_i (1-\xi^2)$
\\\cline{1-2}
{\small Bremsstrahlung}
       &
\multirow{2}{*}{     $ \frac{
     16 n_e^2  G \alpha^2 p_i }{15\omega m_e}\left(
     \left(6 (1+\xi ^2)-\frac{3 (1-\xi ^2)^4+7 (1-\xi^4)^2}{2
   (1+\xi ^2)^3}\right)
{L}+
20 \xi -\frac{6 \xi  (1+\xi ^4)}{(1+\xi ^2)^2}
 +{\cal O}(\xi_s^2)
 \right) 
  $} 
&
\multirow{2}{*}{$L=\log \sqrt{\frac{(1+\xi)^2+\xi _s^2}{(1-\xi )^2+\xi _s^2}}$}
\\
$e e\rightarrow e e\ h$ &&
\\\hline
   \end{tabular}
    \caption{Emission rates of gravitons, $h$, from the indicated process in the non-relativistic limit.   $p_i$ ($p_f$) denotes the initial (final) momentum of the colliding particles in the center-of-mass frame, and $E_i$ is total kinetic energy.
    }
    \label{table:processes}
\end{table*}

The emission rate per unit volume of one graviton in the collision of two particles is given by
\begin{equation}
\frac{d \Gamma}{ \mathrm{d} \omega \mathrm{d} V}\left(1+2 \rightarrow h_\lambda \cdots\right)=\int d n_1 d n_2|{\cal M }(\lambda)|^2{\rm d(  P S)} (2 \pi)^4 \delta^{(4)}\left(p_1+p_2-p-\sum_k p_k\right)\,,
\label{eq:master}
\end{equation}
where $p=(\omega, \bf{p})$ and $\lambda$ are respectively the graviton momentum and helicity, while ${\rm d(  P S)}$ is the phase space associated with the final-state particles,
\begin{equation}
{\rm d(  P S)}=\frac{\omega^2 {\rm d} \Omega_{\bf{p}}}{(2 \pi)^3 2 \omega}   \prod_k \frac{{\rm d}^3 p_k}{(2 \pi)^3 2 E_k}\,.
\label{eq:one}
\end{equation}
In addition, ${\cal M} (\lambda)$ is the graviton emission amplitude, which can be cast as
\begin{equation}
{\cal M}(\lambda)={\cal M}_{\mu \nu} \epsilon^{\mu \nu}_{(\lambda)}(p) \,,
\end{equation}
where polarizations tensors are given by
\begin{align}
 \epsilon^{\mu \nu}_{\pm 2}(p)=\epsilon_{\pm}^\mu(p) \epsilon_{\pm}^\nu(p)\,,
&&\text{with}&&
\epsilon_{\pm}^\mu(p)=\frac{1}{\sqrt{2}}\left(\begin{array}{c}
0 \\
\mp\cos \theta \cos \phi+i \sin \phi \\
\mp\cos \theta \sin \phi-i \cos \phi \\
\pm\sin \theta
\end{array}\right)\,,
&& \text{and}
&& p^\mu=\left(\begin{array}{c}
\omega\\
|\bf{p}|  \sin \theta \cos \phi \\
|\bf{p}| \sin \theta \sin \phi \\
|\bf{p}|  \cos \theta
\end{array}\right)\,.
\end{align}

When perturbation theory applies, ${\cal M}_{\mu \nu}$ in Eq.~\eqref{eq:master} can be calculated from tree-level diagrams, with the vertices obtained from
\begin{equation}
\label{eq:Lfull}
\mathcal{L}=  \frac{1}{2} \partial_\rho h_{\mu \nu} \partial^\rho h^{\mu \nu}-\frac{1}{2} \partial_\rho h \partial^\rho h +\partial_\rho h \partial_\nu h^{\rho \nu}-\partial_\rho h_{\mu \nu} \partial^\nu h^{\mu \rho} 
+\frac{1}{2}\kappa_G\, h_{\mu \nu} T^{\mu \nu} \,,
\end{equation}
where $T^{\mu\nu}$ is the stress-energy-momentum tensor of the Standard-Model particles, and $g_{\mu\nu} = \eta_{\mu\nu} + \kappa_G h_{\mu\nu} $ with $\kappa_G= \sqrt{32\pi G}$. 
The corresponding Feynman rules of relevance for this work are

\begin{tabular}{cc}
\multirow{4}{*}{
\includegraphics[height=0.055\textheight]{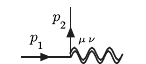}\qquad}& \\
&
$
C^{\rm (fermion)}_{\mu\nu}(p_1,p_2)=\frac{i}{8} \kappa_G\left[2 \eta_{\mu \nu}(\slashed{p}_1+\slashed{p}_2-2 m_f)-(p_1+p_2)_\mu \gamma_\nu-\gamma_\mu(p_1+p_2)_\nu\right]\,, 
$
\\
&\\
&\\
&\\
\end{tabular}
\begin{tabular}{cc}
\multirow{5}{*}{
\qquad\includegraphics[height=0.085\textheight]{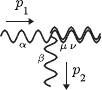}\qquad
}
&\\
&
$
C^{\rm (photon)}_{\alpha\beta\mu\nu}(p_1,p_2)=-\frac{i}{2} \kappa_G\left[ \eta_{\mu \nu} \eta_{\alpha \beta} \, p_1 \cdot p_2-\eta_{\mu \nu} p_{1 \beta} p_{2 \alpha} +\eta_{\nu \alpha} p_{1 \beta} p_{2 \mu}-\eta_{\alpha \beta} p_{1 \nu} p_{2 \mu}
+\eta_{\mu \beta} p_{1 \nu} p_{2 \alpha}\right. 
$\\
&
$
\qquad\qquad\qquad\left.-\eta_{\nu \alpha} \eta_{\mu \beta} p_1 \cdot p_2+\eta_{\nu \beta} p_{1 \mu} p_{2 \alpha} -\eta_{\alpha \beta} p_{1 \mu} p_{2 \nu}+\eta_{\mu \alpha} p_{1 \beta} p_{2 \nu} -\eta_{\nu \beta} \eta_{\mu \alpha} p_1 \cdot p_2  \right] \,,
$\\
&\\
&\\
&\\
\end{tabular}
\begin{tabular}{cc}
\multirow{4}{*}{
\includegraphics[height=0.055\textheight]{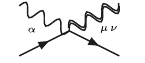}
}&\\
&
\qquad
$
C^{\rm (quartic)}_{\alpha\mu\nu}=-\frac{i}{4} e Z \kappa_G\left[2 \eta_{\mu \nu} \gamma_\alpha-\eta_{\alpha \mu} \gamma_\nu-\eta_{\alpha \nu} \gamma_\mu\right]
\,.
$
\\
&\\
&\\
&\\
\end{tabular}

\noindent
The fermion line is either $e$ or $Z$. These rules agree with existing literature (see e.g.~\cite{Choi:1994ax,Donoghue:1994dn}), and were derived implementing Eq.~\eqref{eq:Lfull} in FeynRules~\cite{Alloul:2013bka}. This process also gives the input needed by CalcHEP~\cite{Belyaev:2012qa} for cross section calculations. Using Package-X~\cite{Patel:2016fam}, we manipulate the symbolic output thus obtained to compute ${\cal M}^{\mu \nu}$.   
For all processes, we verify that 
\begin{equation}
p^\mu {\cal M}_{\mu\nu} =0 \,,
\label{eq:Minvariance}
\end{equation}
as follows from energy-momentum conservation. 

\textbf{Validity of the Born approximation. }
The procedure just described relies on the validity of the Born approximation, which holds true for photoproduction, $\gamma a \to a h $, but becomes problematic for bremsstrahlung processes, $ ab \to ab\,h$.  (Here $a$ or $b$ stands for an electron or a nucleon, as above).
Let us discuss bremsstrahlung first. This issue is closely connected to the perturbativity in the $ab$ collisions, discussed above. This can be understood as follows. The Born approximation assumes that particles in the initial and final states are approximately described by plane waves. While this assumption holds true for gravitons, it does not necessarily apply to electrons or nucleons, as they are subject to mutual interaction. The extent to which a plane wave is a good approximation depends on whether this mutual interaction is perturbative. If it is, the wave function of electrons or nucleons can be expanded perturbatively, resulting in a plane wave with a small additional contribution, whose impact on the scattering rate appears beyond the Born approximation.

The previous argument makes it clear that whenever perturbativity is a problem for graviton bremsstrahlung, it is also a problem for the bremsstrahlung of axions or photons. Interestingly, Weinberg has recently revisited the latter case, providing approximation formulas for photon bremsstrahlung valid beyond the Born approximation~\cite{Weinberg:2019mai} (see also Ref.~\cite{Pradler:2021ppc}).  
Such formulas arise from the non-perturbative wave functions and provide substantial corrections when $v/Z\alpha$ is small. In light of Fig.~\ref{fig:fig1b}, this shows that for the main processes contributing to graviton bremsstrahlung in the Sun --namely, $ee$, $ep$ and $eHe$-- the Born approximation works well except in the last 15\% of the solar radius. There, the temperature is comparatively lower resulting in $v/Z\alpha <1$, invalidating the plane-wave assumption.  
Since bremsstrahlung is dominated by processes at the center, neglecting non-perturbative corrections to the graviton emission is justified. Likewise, this argument applies to axion bremsstrahlung, where accounting for non-perturbative wave functions gives corrections below 10\% (see e.g.~Ref.~\cite{Hoof:2021mld}).

Finally, the aforementioned argument justifies the use of the Born approximation for  photoproduction because 
the charged particles in the initial state, although non-relativistic, do not experience a distortion of their wave function, as in the case of bremsstrahlung. 

\subsection{Photoproduction}

\begin{figure}[b]
\includegraphics[height=0.055\textheight]{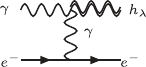}\quad\quad\quad
\includegraphics[height=0.055\textheight]{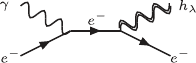}\quad\quad\quad
\includegraphics[height=0.055\textheight]{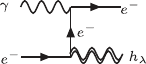}\quad\quad\quad
\includegraphics[height=0.055\textheight]{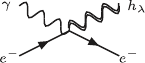}
\caption{Feynman diagrams for photoproduction 
}
\label{fig:photoproduction_feynman_diagrams}
\end{figure}

Having justified the Born approximation, let us note that diagrammatically such collisions correspond to Feynman diagrams where the photon emits $h_{\pm2}$ and then collides with a charged particle (Primakoff-like scattering), or to one photon impinging on a charged particle that subsequently emits $h_{\pm2}$ (Compton-like scattering), or to a diagram involving a contact interaction with all particles.  Using Eq.~\eqref{eq:one}, the amplitude can be cast as
\begin{equation}
{\cal M}_{\alpha\beta} = \overline{u}_3 \Gamma_{\alpha\beta\mu} u_2 \,\epsilon^{\mu}(p_1)\,,
\end{equation} 
where the subscripts $2$ and $3$ refer to the fermion in the initial and final state, respectively, while $1$ refers to the incoming photon.    Following the procedure outlined above, we find 
\begin{eqnarray}
 \label{eq:PhotonProduction}
 \Gamma_{\alpha\beta\mu} =&-& \frac{Z e}{(p_2-p_3)^2} C_{\mu\lambda\alpha\beta}^{\rm(photon)}(p_1,p_3-p_2)  \gamma^\lambda \\
 &+& \frac{Z e}{(p_1+p_2)^2-m_f^2} C^{\rm(fermion)}_{\alpha\beta}(p_1+p_2,p_3) \left(\slashed{p_1}+\slashed{p_2}+m_f\right) \gamma_\mu \nonumber\nonumber\\
 &+&\frac{Z e}{(p_1-p_3)^2-m_f^2} \gamma_\mu  \left(-\slashed{p_1}+\slashed{p_3}+m_f\right)C^{\rm(fermion)}_{\alpha\beta}(p_2,-p_1+p_3)\nonumber\\
  &+& C^{\rm (quartic)}_{\mu\alpha\beta} \,. 
  \nonumber
\end{eqnarray}
In addition to making sure that this satifies Eq.~\eqref{eq:Minvariance}, we also verify gauge invariance by means of the Ward–Takahashi relation
\begin{eqnarray}
p_1^{\mu}\,\left(\overline{u}_3 \Gamma_{\alpha\beta\mu} u_2 \, \epsilon^{\alpha \beta}_{(\lambda)}(p)  \right) =0 \,.  
\label{eq:WT}
\end{eqnarray}
Plugging Eq.~\eqref{eq:PhotonProduction} into Eq.~\eqref{eq:master}, we obtain the photoproduction rates.  Taking the non-relativistic limit, $m_f\gg \omega$, we obtain the differential emission rate 
\begin{align}
 \frac{\mathrm{d}\Gamma}{\mathrm{d}\omega \mathrm{d}V  \mathrm{d}\Omega}\left(\gamma Z\to  Z \,h\right) = n_\gamma n_Z  \delta(\omega-p_i) \, Z^2 \frac{\mathrm{d} \sigma}{\mathrm{d} \Omega}\bigg|_{e}\,, &&\text{with}&&   \frac{\mathrm{d} \sigma}{\mathrm{d} \Omega}\bigg|_{e} =\frac{1}{2} \, \alpha G \, \cot^2\!\frac{\theta}{2} \left(1+\cos^2\theta\right) \,.
 \label{eq:sigma0photoproduction}
\end{align}
This cross section was reported in Ref.~\cite{DeLogi:1977qe, Voronov:1973kga}. As usual, here $\theta$ is the scattering angle in the photoproduction process,  i.e. the angle between the graviton and the incoming particles. 
So far, we have assumed that the photon is massless. However, the surrounding plasma gives photons in the external line an effective mass  given by the plasma frequency, $\omega_{\rm pl}$. A similar modification is required in the photon propagator and vertices. Nevertheless, the solar temperature far exceeds $\omega_{\rm pl}$, allowing us to calculate photoproduction rates by expanding in $\omega_{\rm pl}$. In practice, this approach is effectively equivalent to treating the external photon as massless while incorporating the plasma frequency into the phase space, as reported in the Table~\ref{table:processes} with $\kappa=0$. The resulting cross section agrees with semi-classical calculations~\cite{Galtsov:1974yu} accounting for a non-trivial dispersion (without screening).

As is clear from Eq.~\eqref{eq:sigma0photoproduction}, the total rate diverges due to the pole at $\theta=0$.   The origin of the divergence is the photon propagator in the amplitude of the Primakoff-like diagram, whose denominator $-(p_2-p_3)^2$ equals in the non-relativistic limit the square of the three-momentum transfer, $ |{\bf q}|=2\omega \sin(\theta/2)$.  To gain further intuition on this, let us discuss the case of axions, where a similar divergence occurs for the Primakoff effect. 
As shown in \cite{Raffelt:1985nk}, the divergence is effectively regularized by accounting for correlations among plasma particles. This is closely related to screening, and despite the differences between axions and gravitons, the same remains true for the latter in non-relativistic limit, $m_f\gg \omega$. We repeat the argument here for completeness. 

In the non-relativistic limit,  $|{\bf q}|/m_f \ll 1$, and therefore the charged particles off which the photon scatters may be regarded as at rest during the collision.
As a result,  the photon scatters off a charge distribution, $\rho({\bf  r})$, with a cross section 
\begin{align}\label{eq:ansatz}
	\frac{\mathrm{d} \sigma}{\mathrm{d} \Omega}= \frac{\mathrm{d} \sigma}{\mathrm{d} \Omega}\bigg|_{e}
	\left|F(\mathbf{q})\right|^2 \,,
	&&\text{where}&&
	F({\bf q}) = \frac{1}{e} \int   \mathrm{d} { \bf r} \, e^{i  {\bf  q} \cdot  { \bf r}}  \rho({\bf r}) \,.
\end{align}
The cross section with subscript $e$ refers to electrons as introduced in Eq.~\eqref{eq:sigma0photoproduction}.  For  $N$ point charges $Z_i e$ located at $\mathbf{r}_i$, a explicit computation shows that $F({\bf q}) = \sum_i^N Z_i e^{i {\bf q} \cdot {\bf r}_i }$ as well as
$\left|F(\mathbf{q})\right|^2=\sum_{i}^N Z_i^2+\sum_{\substack{i \neq j}}^N Z_i Z_j \cos \left( \mathbf{q} \cdot \mathbf{r}_{i j}\right)
$
%
where $\mathbf{r}_{i j} \equiv \mathbf{r}_i-\mathbf{r}_j$. This must be averaged accounting for the statistical ensemble describing the plasma. Clearly, the first term is associated with  individual charges, while  the second one describes interference effects. The latter averages to zero for very diluted plasmas ($\kappa\to 0$) because their particles are uncorrelated. In that case, the total rate in Eq.~\eqref{eq:ansatz} simply consists of adding the contribution of individual charges. Nevertheless, in real plasmas particles mutually interact through their Coulomb fields and their motion is slightly correlated. Concretely, the potential, $\phi_i$,  around a given particle is sourced by a charge cloud with density
\begin{align}
\rho_i(r)=Z_i e\left(\delta^3(\mathbf{r})-\frac{\kappa^2}{4 \pi} \frac{e^{-\kappa r}}{r}\right) = -\nabla^2 \phi_i\,.
\label{eq:phii}
\end{align}%
That is, the Coulomb field of any charge is screened over distances larger than about $\kappa^{-1}$. 
To determine the aforementioned average, and hence the effect of correlations, we note that the   potential energy associated with Eq.~\eqref{eq:phii} is given by $U=\frac{1}{2} e\sum^N_{i\neq j}Z_i \phi_i $. Employing the Boltzmann factor associated with this potential energy, $e^{-U/T}$, at leading order in $\alpha$  one finds that thermal average is
\begin{align}
\sum^N_{j\neq i} Z_j \langle  \cos \left( \mathbf{q} \cdot \mathbf{r}_{i j} \right) \rangle = - \frac{ Z_i \kappa^2}{\kappa^2+|{\bf q}|^2} \,,
&&
\text{which leads to}
&&
\left\langle\left|F(\mathbf{q})\right|^2\right\rangle=\sum_{i=1}^N Z_i^2\left(1-\frac{\kappa^2}{\kappa^2+|\mathbf{q}|^2}\right)\,.
\label{eq:corr}
\end{align}
Hence, Eq.~\eqref{eq:ansatz} implies that~\cite{Raffelt:1985nk}
\begin{align}
	\left\langle\frac{\mathrm{d} \sigma}{\mathrm{d} \Omega}\right\rangle = \sum_{i=1}^N
	Z_i^2  \frac{\mathrm{d} \sigma}{\mathrm{d} \Omega}\bigg|_{e}  \frac{|\mathbf{q}|^2}{\kappa^2+|\mathbf{q}|^2} \,.
	\label{eq:Raffelt}
\end{align}
In summary, one must add individual cross sections introducing a global factor determined by the screening scale. This clearly regularizes the divergence at $|{\bf q}|=0$. Following this procedure, we obtain the rates reported in Table~\ref{table:processes}.  A few comments are in order.

While the Compton process for axions arises from a renormalizable coupling, $g_{aee}$, leading to a cross section independent of the momentum transfer in the non-relativistic limit, the Primakoff process results from a non-renormalizable coupling, $g_{a\gamma\gamma}$, and leads to a differential cross section scaling as the fourth power of the momentum transfer~\cite{Raffelt:1985nk}. Thus, only the latter exhibits a divergence, raising the question of whether the Compton-like and Primakoff-like diagrams for gravitons must be separated somehow.  Nonetheless,  unlike photoproduction of axions, where such a separation  is feasible, for gravitons that cannot be done in a meaningful manner. First, because they depend on the same coupling, namely, the gravitational coupling in Eq.~\eqref{eq:Lfull}. Moreover, although the identity in Eq.~\eqref{eq:WT} is separately satisfied by the Compton-like diagrams of Fig.~\ref{fig:photoproduction_feynman_diagrams},  that is not the case of the identity associated with energy-momentum conservation, Eq.~\eqref{eq:Minvariance}.  In the light of this, we apply Eq.~\eqref{eq:Raffelt} to photoproduction of gravitons as a whole. 

One might ignore the correlations given by Eq.~\eqref{eq:corr} and assume that the incoming photon scatters off a single charge distribution given by Eq.~\eqref{eq:phii}. This is the regularization advocated 
by Ref.~\cite{DeLogi:1977qe}. The resulting cross section is that in Eq.~\eqref{eq:Raffelt} with $|\mathbf{q}|^2/(\kappa^2+|\mathbf{q}|^2) \to |\mathbf{q}|^4/(\kappa^2+|\mathbf{q}|^2)^2$.
As explained in the main text, this issue has been revisited recently for the Primakoff process~\cite{Hoof:2021mld}, accounting for the velocity in the form factor. Corrections to Eq.~\eqref{eq:Raffelt} have been deemed negligible. Coupled with this, let us remark that, as opposed to bremsstrahlung or two-two scattering, the effect of screening here cannot be understood as an effective mass for the photon for several reasons. First, the effective mass of transverse photons in the external line is given by the plasma frequency, $\omega_{\rm pl}$, and not by the Debye-Hückel screening scale, $\kappa$, which in the Sun satisfies $\omega_{\rm pl} \ll \kappa$. See Fig.~2 of the main text. In fact, even if one adopts an ad-hoc prescription in which the virtual photon in Fig.~\ref{fig:photoproduction_feynman_diagrams} has a mass $\kappa$, while the incoming external photon has a mass $\omega_{\rm pl}$, one quickly finds inconsistencies. For instance, this approach spoils Eq.~\eqref{eq:Minvariance}.

\subsection{Bremsstrahlung}

\begin{figure}[b]

\includegraphics[height=0.08\textheight]{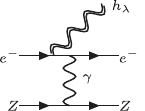}
\includegraphics[height=0.08\textheight]{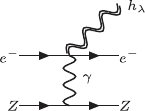}
\includegraphics[height=0.08\textheight]{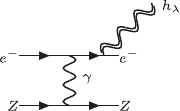}
\includegraphics[height=0.08\textheight]{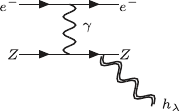}
\vspace{1cm}
\includegraphics[height=0.08\textheight]{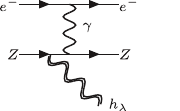}
\includegraphics[height=0.08\textheight]{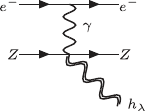}
\includegraphics[height=0.08\textheight]{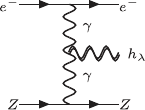}

\caption{ Feynman diagrams for bremsstrahlung in $eZ$ collisions.
}
\label{fig:bremsstrahlung_feynman_diagrams}
\end{figure}

\begin{figure}[t]
\includegraphics[height=0.38\textheight]{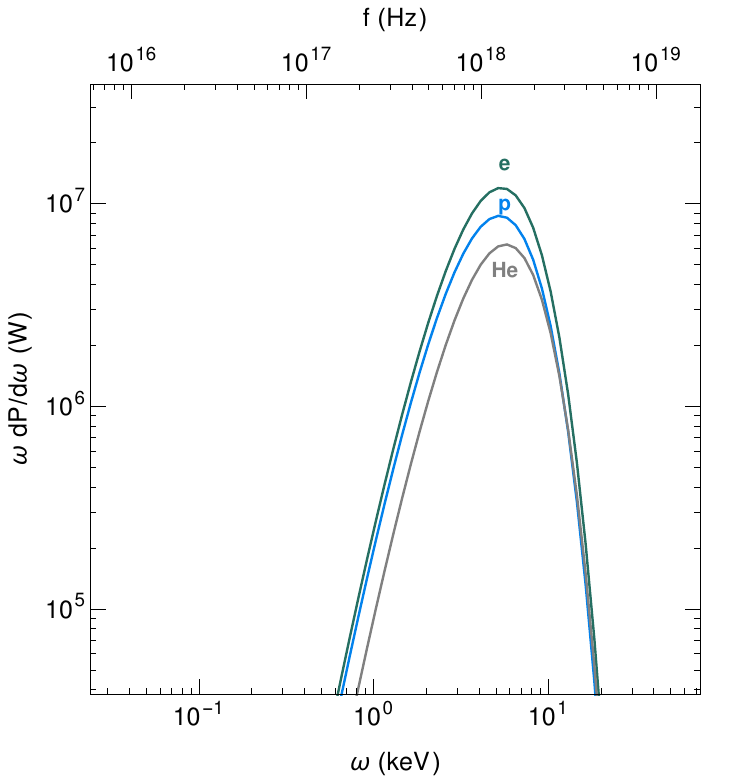}
\includegraphics[height=0.38\textheight]{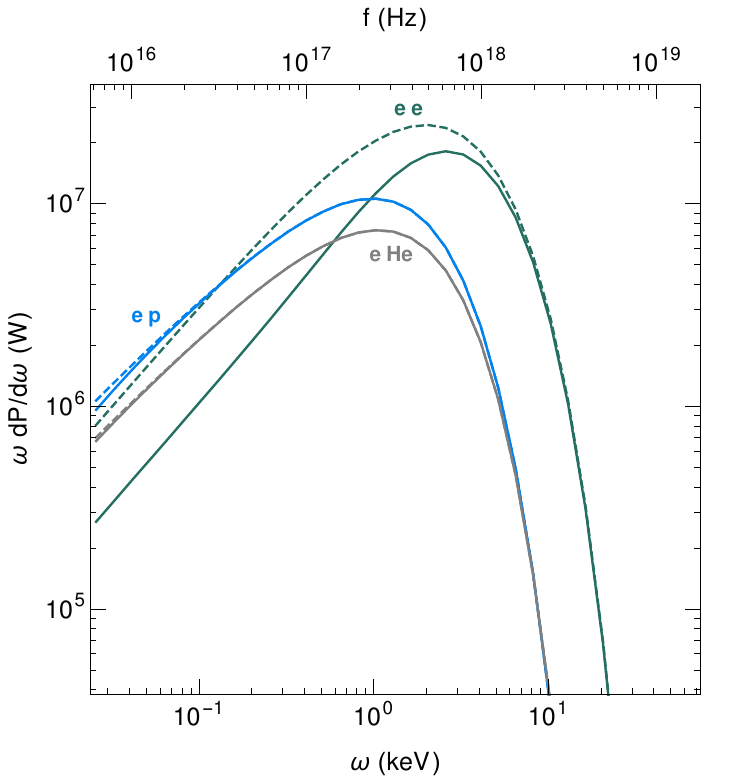}

\caption{ 
\emph{Left:} Contribution of photoproduction processes to the solar GW spectrum.  
\emph{Right:} Contribution of bremsstrahlung processes to the solar GW spectrum. Solid lines are the full results determined by Eqs.~\eqref{eq:apeZ} and ~\eqref{eq:apee}. The dashed lines are the leading results in the screening scale, reported in Table~\ref{table:processes}. 
}
\label{fig:soft}
\end{figure}

In the Born approximation, there are seven Feynman diagrams associated with bremsstrahlung in $eZ$ collisions, see Fig.~\ref{fig:bremsstrahlung_feynman_diagrams}. For $ee$ collisions, additional diagrams are obtained by symmetrizing the final state of those shown in Fig.~\ref{fig:bremsstrahlung_feynman_diagrams}. Note that bremsstrahlung from two-nucleon states are negligible. 
To calculate the corresponding $M^{\mu\nu}$, we follow the procedure outlined above. Concretely,  we manipulate the symbolic output provided by CalcHEP~\cite{Belyaev:2012qa}  using Package-X~\cite{Patel:2016fam}. Crucially,  we verify that Eq.~\eqref{eq:Minvariance} is satisfied. Subsequently, we use this together with the appropriate polarization tensors to compute ${\cal M}(\lambda)$.  To account for screening, we modify the photon propagators with an effective mass equal to the screening scale. As argued above, that is the correct prescription for slow particles such as the non-relativistic plasma components (in contrast to the external photon in the photoproduction process described above). From this, we calculate the differential cross sections in energy. 
To the best of our knowledge, the expression for these have never been reported in the literature in a general form. This is not surprising as the aforementioned procedure is a daunting task, even in the the non-relativistic limit. After algebraic manipulation, in such a limit we find 
\begin{eqnarray}
\label{eq:apeZ}
\frac{{\mathrm d}\sigma\!v }{{\mathrm d} \omega} \Bigg|_{eZ\to eZ\, h}&=&\frac{32 Z^2 \alpha ^2 G p_i}{15 \omega}   \left(\frac{1}{m_e}+\frac{1}{m_Z} \right) \left[3 (1+\xi ^2+\xi _s^2) L \right. \\
&&\left.+10 \xi +\frac{1}{3} \xi _s^2\left(\frac{(1-\xi )^2 \left(18 (1+\xi)^4+29 (1+\xi)^2 \xi _s^2+12 \xi _s^4\right)}{\left((1+\xi)^2+\xi _s^2\right)^3}-(\xi\to-\xi)\right) \right]  \,,\nonumber
\end{eqnarray}
with notation $L$, $\xi$ and $\xi_s$ introduced in Table~\ref{table:processes}. In this calculation we do not take the limit $m_Z\gg m_e$, we simply assume that both nucleons and electrons are non-relativistic. Likewise
\begin{eqnarray}
\label{eq:apee}
&&\frac{{\mathrm d}\sigma\!v }{{\mathrm d} \omega}  \Bigg|_{ee\to ee\, h}=
\frac{32 \alpha ^2 G p_i}{15 \omega m_e} \left\{ 20 \xi -\frac{6 \xi 
   (1+\xi ^4)}{(1+\xi ^2)^2}+ \left[6 (1+\xi ^2)-\frac{3 (1-\xi ^2)^4+7 (1-\xi ^4)^2}{2 (1+\xi ^2)^3}\right.\right.\\
&&\left.\left. \frac{1}{2} \xi_s^2
  \left(\frac{6 (\xi ^4+1) (1-\xi ^2)^2}{(\xi ^2+1)^2(\xi ^2+\xi_s^2+1)^2}+\frac{2 (\xi ^4-4 \xi
   ^2+1) (1-\xi ^2)^2}{(\xi ^2+1) (\xi ^2+\xi_s^2+1)^3}+\frac{13 \xi ^8+22 \xi ^4+13}{(\xi
   ^2+1)^3 (\xi ^2+\xi_s^2+1)}+15\right)\right]L
 \right.\nonumber\\
 && + \left. \xi_s^2 \left[2 (1-\xi ^2)^2 \left(-\frac{(1-\xi )^4-80 \xi ^2}{16\, \xi^2 (\xi +1)^2 ((\xi +1)^2+\xi_s^2)}-\frac{3 (\xi ^4+1)
   \xi_s^2+4 (\xi ^6+1)}{8 \xi  (\xi ^2+1)^2 (\xi ^2+\xi_s^2+1)^2}+\frac{6 (\xi +1)^2+5 \xi_s^2}{3 ((\xi +1)^2+\xi_s^2)^3}\right)\right.\right.\nonumber\\
&&\left.\left.-(\xi\to-\xi) \right] \right\} \nonumber\,.
\end{eqnarray}  
For $\xi_s=0$, these cross sections agree fully with those reported by Gould in Refs.~\cite{Gould:1981an,Gould:1985}.  The total emission rate is given by
\begin{align}
\omega\frac{\mathrm{d}\Gamma}{\mathrm{d}\omega \mathrm{d}V}  \Bigg|_{\rm Bremsstrahlung} = \frac{1}{2}n_e^2 \frac{{\mathrm d}\sigma\!v }{{\mathrm d} \omega}  \Bigg|_{ee\to ee\, h} + \sum_Z n_e n_Z  \frac{{\mathrm d}\sigma\!v }{{\mathrm d} \omega} \Bigg|_{eZ\to eZ\, h}\,.
\label{eq:ratestocross}
\end{align}
 The partial contributions of each channel are shown in Fig.~\ref{fig:soft}.
To leading order in the screening scale the resulting expressions are reported in Table~\ref{table:processes}. While these approximations are quite accurate for $eZ$ bremsstrahlung, they are less precise for $ee$ bremsstrahlung, as shown in Fig.~\ref{fig:soft}. This underscores the critical importance of incorporating screening effects in the bremsstrahlung rates.

\textbf{Soft emission.} Let us consider the initial state $eZ$. A soft graviton leads to $\xi\to1$. Taking $\xi_s \to 0$, the effect of screening appears only in the logarithm with $L\to L_{\rm Coulomb}$, and the corresponding emission rate can be cast as

\begin{align}
\omega\frac{\mathrm{d}\Gamma}{\mathrm{d}\omega \mathrm{d}V}  \Bigg|_{eZ\to eZ\, h}=\frac{64 \alpha^2  Z^2 G v }{5}  L_{\rm Coulomb}n_e n_Z 
=\frac{8G }{5 \pi}  n_e n_Z    \, \mu^2 v^5 \sigma_V^{(eZ)}\,.
\end{align}
In the last equation, we use the viscosity cross section obtained in Eq.~\eqref{eq:approx} with $E_i =\mu v^2/2$. Note that $v= p_i \left(\frac{1}{m_e}+\frac{1}{m_Z}\right)$.  Similarly, for $ee$ Eq.~\eqref{eq:apee} reduces to 

\begin{align}
\omega\frac{\mathrm{d}\Gamma}{\mathrm{d}\omega \mathrm{d}V}  \Bigg|_{ee\to ee\, h}=\frac{32 \alpha^2  Z^2 G v }{5}  L_{\rm Coulomb}n_e^2  
=\frac{4G }{5 \pi}  n_e^2    \, \mu^2 v^5 \sigma_V^{(ee)}\,.
\end{align} 
Here $v= 2p_i /m_e$. As explained in the text,  this was first proven by Weinberg~\cite{Weinberg:1965nx}, who used it to estimate the total power emitted by the Sun.

\section{Bound-bound and free-bound transitions}

Bremsstrahlung can be considered a free-free transition because unbound charged particles emit radiation as they scatter. One may also consider free-bound and bound-bound transitions. In the free-bound case, two charged particles, such as a proton and an electron, recombine to form a hydrogen atom, emitting a graviton. In the bound-bound case, both the initial and final states are atomic hydrogen, and the GW emission arises from transitions between atomic levels. The motivation for considering these processes is that when the radiated particle is an axion, they significantly contribute to axion emission from the Sun. In fact, in some solar models, spectral features from bound-bound and free-bound transitions stand out, offering distinct signatures in the solar axion spectrum. Given the close analogy between axions and GWs, one might expect a similar effect for GWs. However, the purpose of this appendix is to show that, due to the quadrupolar nature of graviton emission, free-bound and bound-bound transitions are completely subdominant.

\textbf{Free-bound transitions.} In analogy to Eq.~\eqref{eq:ratestocross}, we have

\begin{align}
\frac{\mathrm{d}\Gamma}{\mathrm{d}\omega \mathrm{d}V}  \Bigg|_{\rm Free-Bound} = \sum_Z n_e n_Z  \frac{ \langle{\mathrm d} \sigma v  \rangle }{{\mathrm d} \omega} \Bigg|_{eZ\to Z^*h}\,.
\label{eq:ratetocross_freebound}
\end{align} 
In this equation, $Z^*$ represents a bound state of an electron and the nucleus of charge $Z$. The cross section appearing here is known as recombination in the literature and is given by
\begin{align}
 \frac{{\mathrm d}\sigma\!v }{{\mathrm d} \omega} \Bigg|_{eZ\to Z^*h}=\frac{2 G\omega^5}{5} \left(D_{ij}^* D_{ij} -\frac{1}{3} |D_{ii}|^2 \right) \,.
\end{align}
where $D_{ij}$ is the quadrupole moment of the atomic transition
\begin{align}
 D_{ij} = m_e\int \psi_{b}^* ({\bf x} ) x_i x_j \psi_a({\bf x  })  d^3x \,.
\label{eq:Dformula}
\end{align}
The atomic wave function for the final state is normalized as $\int | \psi({\bf x}) |^2 d^3x = 1$, while the initial state wave functions are taken as scattering states with asymptotic plane-wave behavior. 

\begin{figure}[t]
\includegraphics[height=0.38\textheight]{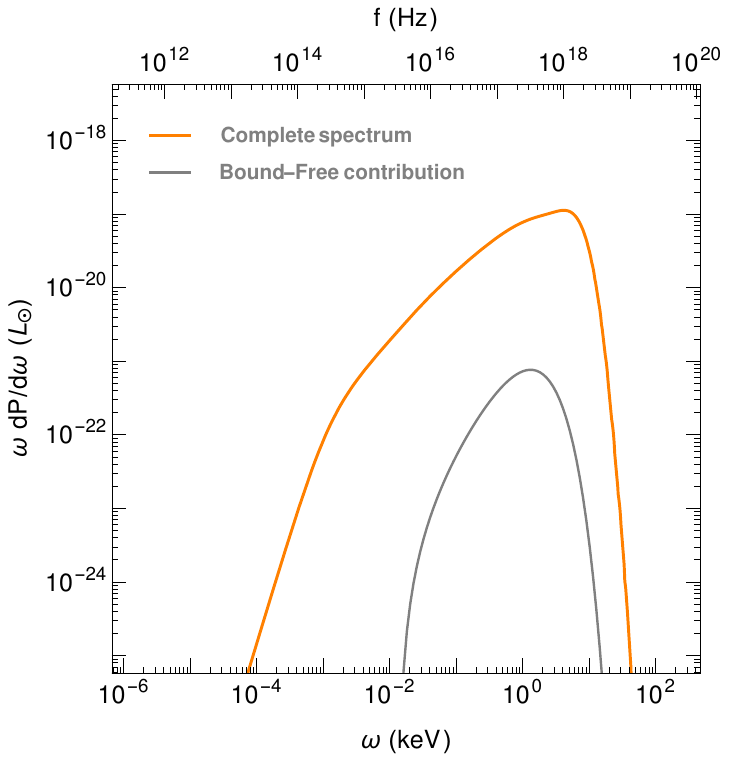}
\caption{ 
 Contribution of the free-bound transition into the fundamental state of hydrogen to the GW spectrum of the Sun. See text for details
}
\label{fig:free-bound}
\end{figure}

To estimate this contribution, we adopt the following approximations. We work in the Born regime and neglect screening. This means that we calculate recombination rates assuming plane waves for the incoming particles in the initial state, which then form a bound state via Coulomb interaction. Accounting for screening will simply suppress the rates, while the Born approximation is justified in roughly the innermost 85\% of the Sun, as argued for free-free transitions. Under these assumptions, we find that the cross section for recombination into the fundamental state of hydrogen-like systems is given by
\begin{equation}
 \frac{{\mathrm d}\sigma }{{\mathrm d} \omega}  \Bigg|_{eZ\to Z^*h} \approx \left(\frac{2\omega^2}{p_i^2} \right) \frac{3\times 2^{10}\pi}{5} \frac{p_i^5 a_B^5}{\left(1+p_i^2 a_B^2\right)^5}G \, \delta\left(\omega-I -\frac{p_i^2}{2m_e}\right) \,,
 \label{eq:crossFB}
\end{equation}
where $a_B=1/Z\alpha m_e$ is the Bohr radius and $I = \alpha^2 Z^2 m_e/2$ is the energy of ionization. 
As a check, we note that, due to detailed balance~\cite{berestetskii1982quantum}, omitting the factor in parentheses yields the cross section for graviton-induced ionization,  $h Z^* \to eZ$, that is, the cross section of the inverse process. This matches previously found results for hydrogen~\cite{Boughn:2006st, Rothman:2006fp}. The thermal average appearing in Eq.~\eqref{eq:ratetocross_freebound} can be readily determined as  
\begin{align}
 \frac{ {\mathrm d}\langle\sigma v \rangle }{{\mathrm d} \omega}  \Bigg|_{eZ\to Z^*h} \approx \frac{1536 \sqrt{\pi } \alpha ^5 Z^5   G m_e e^{-\omega/T}}{5 T^{3/2} \sqrt{\omega }}     \,,
\end{align}
where for simplicity we assume $\omega \gg I$ (or equivalently  $p_i a_B \gg 1$). Using this and following  the procedure detailed in the main text,  we find the contribution of free-bound transitions into the ground state of hydrogen to the GW spectrum of the Sun, as shown in Fig.~\ref{fig:free-bound}.
We note that the spectrum for free-bound emission is continuous starting from the ionization energy, and as is evident from the figure, its contribution is subdominant.
This result can be understood from the following estimate

\begin{align}
 \frac{\left\langle \frac{\mathrm{d}\Gamma}{\mathrm{d}\omega \mathrm{d}V} \right\rangle \Big|_{\rm Free-Bound}}{
 \left\langle \frac{\mathrm{d}\Gamma}{\mathrm{d}\omega \mathrm{d}V} \right\rangle \Big|_{\rm Bremsstrahlung}} \sim \frac{16 \sqrt{3 \pi }  }{ L_{\rm Coulomb} } Z^3 \alpha ^3 \left(\frac{m_e}{T}\right)^{\frac{3}{2}} \left(\frac{\omega}{T}\right)^{\frac{1}{2}} e^{-\omega/T} \,. 
 \label{eq:ratioFBtoFF}
 \end{align}
 
For example, for hydrogen at $\omega \sim T\sim 1$\, keV, this gives a ratio of approximately $10^{-2}$, in precise agreement with Fig.~\ref{fig:free-bound} after accounting for the branching ratio of $ep$ bremsstrahlung. This small factor stems from the third power of $\alpha$ entering Eq.~\eqref{eq:ratioFBtoFF}, resulting from 
the quadrupole nature of the transition. This sharply contrasts with the case of axions, where free-free and free-bound transitions are comparable~\cite{Redondo:2013wwa}.  Accounting for screening would further suppress this ratio.

 \textbf{Bound-bound transitions:} The graviton resulting from bound-bound transitions, that is, from the decay of excited atomic states, is given by
\begin{align}
 \left\langle \frac{\mathrm{d}\Gamma}{\mathrm{d}\omega \mathrm{d}V} \right\rangle \Bigg|_{\rm Bound-Bound} = \sum_{a}  n_{a}\sum_{b}   \Gamma \Big|_{a\to b h} \phi_\omega \,. 
\label{eq:ratetocross_boundbound}
\end{align} 
Here, $n_{a}$ represents the densities of atoms in state $a$, 
while the function $\phi_\omega$ is sharply peaked at $\omega =E_a-E_b$ and satisfies $ \int \mathrm{d} \omega  \phi_\omega =1$. This function accounts for the broadening of the emission lines, which may be intrinsic or due to Doppler shifts arising from the thermal distribution of atoms. Additionally, the decay rate appearing in Eq.~\eqref{eq:ratetocross_boundbound} is given by~\cite{Weinberg:1972kfs}
\begin{align}
 \Gamma \Big|_{a\to b h} =\frac{2 G\omega^5}{5} \left(D_{ij}^* D_{ij} -\frac{1}{3} |D_{ii}|^2 \right) \,.
\end{align}
Here, $D_{ij}$ denotes the quadrupole moment of the atomic transition, computed using Eq.~\eqref{eq:Dformula}, with both initial and final state wave functions of the atomic states, normalized such that $\int  | \psi({\bf x}) |^2 d^3x =1$. Being completely analogous to ordinary E2 transitions, the same selection rules apply; in particular, only atomic states differing by two units of angular momentum can induce these transitions. For hydrogen, the simplest of these transitions is $3d\to1s$.  Moreover, for hydrogen-like systems, on dimensional grounds, $D_{ij} \sim m_e a_B^2$. Hence,  $\Gamma \sim \frac{2}{5} G \omega^5 m_e^2 a_B^4 \sim \frac{1}{80} Z^6 \alpha^6 G m_e^3$, where we take the graviton frequency on the order of typical energy levels, $\omega\sim \frac{1}{2} Z^2 \alpha^2 m_e$. Then, the contribution of one individual transition in Eq.~\eqref{eq:ratetocross_boundbound} compared to Bremsstrahlung is of order 

 \begin{align}
 \frac{\left\langle \frac{\mathrm{d}\Gamma}{\mathrm{d}\omega \mathrm{d}V} \right\rangle \Big|_{\rm Bound-Bound}}{
 \left\langle \frac{\mathrm{d}\Gamma}{\mathrm{d}\omega \mathrm{d}V} \right\rangle \Big|_{\rm Bremsstrahlung}} \sim  \frac{ \frac{1}{80} n_{a}  Z^6 \alpha^6 G m_e^3 \,\omega\, \phi_\omega}{ \frac{32}{5} n_e n_Z \alpha^2 Z^2 G (T/m_e)^{1/2} L_{\rm Coulomb} } \sim  \frac{ Z^4\alpha^4}{512 \sqrt{2\pi}L_{\rm Coulomb} }\left[\frac{n_a (m_e T)^{\frac{3}{2}} }{n_e n_Z} \right]  \left( \frac{m_e }{T}\right)^2\!\! \left( \frac{m_a}{T}\right)^{\frac{1}{2}}\,. 
 \label{eq:ratioBBtoFF}
 \end{align}
 For this estimate we use the fact that $\omega \phi_\omega \big|_{\omega=E_b-E_a} \simeq (m_a/2\pi T)^{1/2}  $, which results from neglecting the intrinsic width and employing the Maxwell-Boltzmann distribution. Furthermore,
the Saha equation establishes that $n_a/(n_Z n_e) = (m_e T/2\pi)^{-3/2} U $, where $U$ is a factor smaller than one determining the ionization fraction.   Using this, the ratio in Eq.~\eqref{eq:ratioBBtoFF} gives approximately $10^{-4} U Z^4$. For the case of hydrogen, $U$ is smaller than $10^{-4}$ in the solar core, see  e.g.~\cite{2025SoPh..300....3B}. Therefore, the bound-bound transitions are completely subdominant with respect to bremsstrahlung, even taking $Z$ very large. This sharply contrasts with the case of axions, where free-free and bound-bound transitions are comparable~\cite{Redondo:2013wwa}. As we have seen, this difference arises from the large powers of $\alpha$ in Eq.~\eqref{eq:ratioBBtoFF}, which ultimately stem from the quadrupole nature of the transition.

 We once again highlight the distinction with respect to axions, where free-free, free-bound, and bound-bound dipolar transitions contribute comparably to the emission rates.


\end{document}